# Elliptic Flow of Particles under the Influence of Electromagnetic Field Evolution in Relativistic Heavy Ion Collision

Tewodros Gezhagn[1, 2*], A. K. Chaubey[1]

[1]Department of Physics, Addis Ababa University, P.O. Box 1176, Addis Ababa, Ethiopia
[2]Department of Physics, Aksum University, P.O. Box 7080, Aksum, Ethiopia

**Abstract**     **Review Article**

The bending of flow of identified particles from relativistic heavy ion collision is investigated using the iEBE-VISHNU framework. The Maxwell's equations are applied to compute the incremental drift velocity and the change in elliptic flow of particles from the four sources of the electric force, which are of coulomb ($E_C$), Lorentz ($E_L$), Faraday ($E_F$) and Plasma based. We find out that the field evolution arouse flow at lower transverse momentum and suppress it at higher. Heavier particles get higher initial push, and particles and their anti-particles get crudely same elliptic flow changes. More over, elliptic flow is found to show different percentage increase for different collision energies. To conclude, the present study shows that beside the inclusion of electromagnetic field, the increase in collision energy affects the elliptic flow of particles in non uniform fashion through out the evolution. Further study by softening many of the crude assumptions we made and keeping the functionality of parameters is needed to establish a better understanding on the electromagnetic field evolution and its effects on the created system.

# INTRODUCTION

Quenching our burning desire in understanding the very physics of nature happen to look on the hand of High Energy Nuclear Collision Physics. Assuming that the Big Bang actually happened, in the first moments after it, the universe was extremely hot and dense. As the universe cooled, conditions became just right to give rise to the building blocks of matter: the quarks and electrons of which we are all made. A few millionths of a second later, quarks aggregated to produce protons and neutrons. Within minutes, these protons and neutrons combined into nuclei. As the universe continued to expand and cool, things began to happen more slowly. It took 380,000 years for electrons to be trapped in orbits around nuclei, forming the first atoms [1].

Heavy-ions are extended objects and the system created in a head-on collision is different from that in a peripheral collision. To study the properties of the created system, collisions are therefore categorized by their centrality. Theoretically the centrality is defined by the impact parameter b (see Fig-1) which, however, cannot be directly observed. Experimentally, the collision centrality can be inferred from the measured particle multiplicities, given the assumption that the multiplicity is a monotonic function of b. Ultra-relativistic heavy-ion collisions at RHIC and the LHC produce fireballs made of extraordinarily hot matter. Due to enormous pressure gradients between the fireball center and the surrounding vacuum, these fireballs undergo explosive collective expansion, cooling down rapidly through several different states of matter, finally fragmenting into thousands of free-streaming hadrons whose energy and momentum distributions can be detected in the detectors set up around the collider rings. The evolution history of these ``Little Bangs'' has much similarity with the Big Bang that created our Universe. Although they differ dramatically in geometric size and are governed by different fundamental physics (gravity for the big bang, but strong interaction for the little bang), the dynamical evolution of the little bangs share a lot of similarities with the big bang [2].

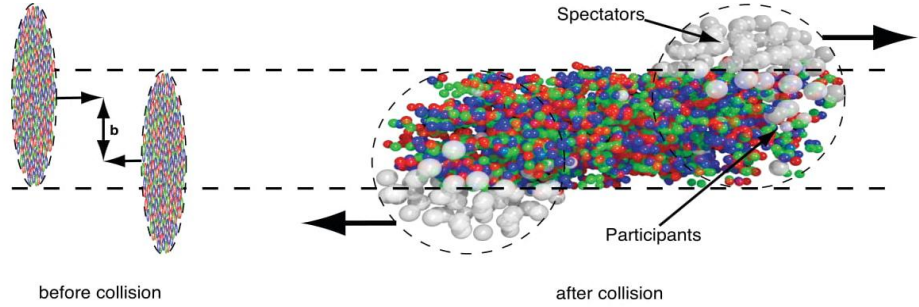

FIG. 1: Left: The two heavy-ions before collision with impact parameter **b**. Right: The spectators continue unaffected, while in the participant zone particle production takes place.

The theory toolkit in relativistic heavy ion physics is quite diverse. It includes QCD perturbation theory in the vacuum and in a thermal medium (especially for the description of jets and heavy quarkonia); semiclassical gauge theory (for the description of the initial conditions reached in the nuclear collision); lattice gauge theory (for static thermodynamic properties of QCD matter, such as its equation of state and color screening); holographic methods mapping strongly coupled gauge theories on their gravity duals (for transport properties and the dynamics of thermalization); and transport theory, especially viscous hydrodynamics (for the evolution of the bulk matter) [3]. Possible signals and probes for the quark-gluon plasma have been investigated for around 30 years since the birth of the field. Such signatures include: collective flow [4, 7], strangeness enhancement [8, 9], charmonium suppression [10, 11], thermal photon and dilepton emission [12, 13], jet quenching [14, 15], critical fluctuations [16], and others. Some of the predicted signature were already found in earlier heavy ion experiments at AGS and SPS energies [17].

However, none of these signatures allow individually to prove QGP formation, as they are contaminated by the dynamical evolution of the fireball through various stages, usually from the very early pre-equilibrium stage through (perhaps) a QGP phase to the late hadronic stage. The combination of three observations at RHIC; the measurements of strong anisotropic collective flow, valence quark number scaling of the elliptic flow $v_2$, and jet quenching finally convinced the community that the QGP has been successfully created [17, 18]. This led to the announcement in 2005 that the QGP had been discovered at RHIC [17, 19]. Over the past decade, relativistic hydrodynamics has established itself as an indispensable component in modeling the collective dynamics of the quark-gluon plasma (QGP) produced in relativistic heavy ion collisions [20]. Over the past decade, relativistic hydrodynamics has established itself as an indispensable component in modeling the collective dynamics of the quark-gluon plasma (QGP) produced in relativistic heavy ion collisions [21-23].

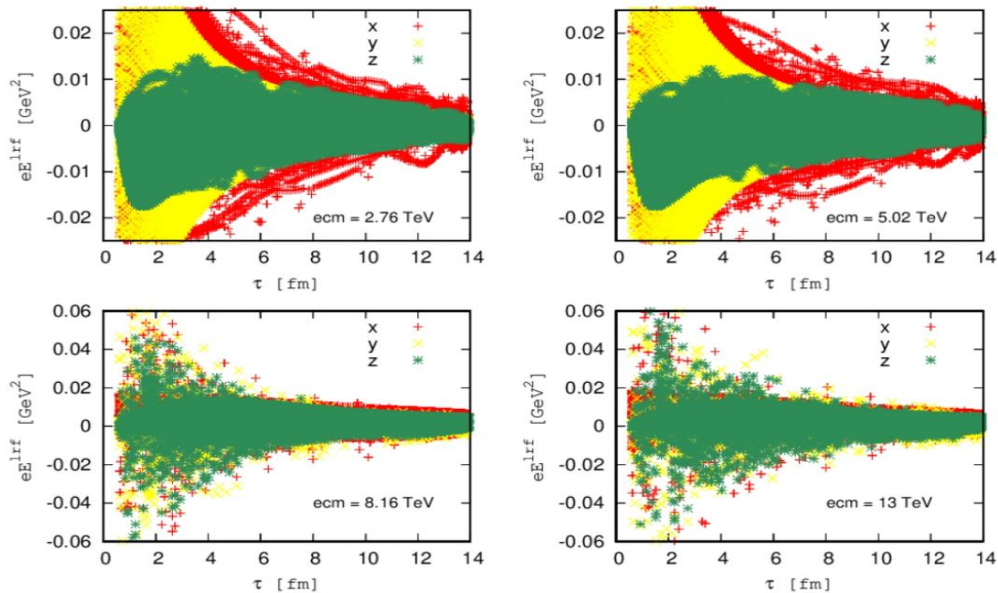

FIG. 1: (Color online) Illustration of the evolution of the electromagnetic field created in relativistic heavy-ion collision after a Pb+Pb collision with 20-30 centrality (impact parameters in the range 6.24 fm $< b <$ 9.05 fm), rapidity window of $|\eta| < 4$ and with a collision energy $\sqrt{s} = 2.76$ A TeV, 5.02 A TeV, 8.16 A TeV and 13 A TeV. shows the three components of the electric field in the local fluid rest frame at points on the freeze-out surface.

Studying the hydrodynamics evolution which is affected with different parameters like, the starting time, initial energy profile, initial flow velocity and viscous stress tensor, The equation of state (EOS), the specific shear viscosity, the kinetic decoupling temperature and many more [20]. As far as Heavy Ion Nuclear Collision is concerned, we are dealing with moving charges, whether the participants, spectators or just the QGP is in mind; which leads us to get concerned of the Electromagnetic fields created from those charges. A huge ongoing effort is being employed in producing a full Magnetohydrodynamics analysis and part of it is understanding how the electromagnetic fields arise and its effect on the evolution. Our work totally deals with understanding how these fields are created and how they affect the hydrodynamic Flow of those elementary particles created in heavy ion collision in relation with other observables.

From many of the parameters affecting flow (starting time, initial energy profile, initial flow velocity and viscous stress tensor, equation of state (EOS), specific shear viscosity, kinetic decoupling temperature and many more [20], we picked the electromagnetic field whose effect is not widely considered by many of the hydro models. In order to address this problem, our model set-up followed three basic steps; first comes simulating the dynamical evolution of the medium produced in Pb-Pb collision at ecm = 2.76 TeV using the iEBE-VISHNU frame-work, compute the drift velocity from the electromagnetic field applying the Maxwell's equation alone using the information we have from the hydrodynamic simulator in order to add this with the velocity of the dynamical system and finally find out the flow harmonics by injecting the total velocity deduced to the particle sampler back in the framework. Now let's first explore where from the electric fields arise.

**The Elecro-Magnetic Field**
The possible origins of the created electric force are the flying spectators, the plasma itself and other two, of magnetic field origin. From the positively charged spectators that fly away the collision zone, there exist an electric force on the charged plasma produced; this is a direct coulomb sourced field. secondly, from the spectator nucleons and from the charge density deposited in the plasma by the nucleons participating in the collision, we have a non-vanishing outward-pointing component of the electric field already in the lab frame. thirdly, the hydrodynamic fluid exhibits a strong longitudinal flow velocity along the beam direction in the direction of which we have coulomb field; and because of the perpendicular magnetic field B created in the system there exist Lorentz force that can contribute to the directional pushes particles experience. Finally, there is an induced electric current because of the change in magnetic field Faraday. A more detailed explanation about the four sources of electric forces coulomb ($E_C$), Lorentz ($E_L$), Faraday ($E_F$) and Plasma based can be found in [24]. Inorder one compute the incremental drift velocity $v_{drift}$ caused by the electromagnetic forces, the electric (E) and magnetic fields (B) evolved should be known from the governing, Maxwell's equation.

**The Model Setup**
A nucleus-nucleus collision at relativistic energy passes through different stages [25], and there is no unique theoretical tool to describe the whole heavy ion collision process from the very beginning till the end and this forces as set up a model. We have used a similar model-set up as in [24] except we extend the drift velocity for more charge types and evaluate the effects for different particles. Following the set-up, we simulate the dynamical evolution of the medium produced in Pb-Pb collision at ecm = 2.76, 5.02, 8.16 and 13 A TeV using the iEBE-VISHNU frame-work and the electromagnetic fields evolution which is computed alone. In general, according to the model set up, there are three basic jobs to be done. The first is simulating the dynamical evolution using same Monte-Carlo–Glauber initialization. Second one is compute the additional velocity induced by the electromagnetic field alone and add this with velocity from the dynamical system itself. In the last step we injected this final velocity to the particle sampler in the framework.

**The iEBE-VISHNU frame-work**
In this hybrid package, there is a specific code simulating each stage of the evolution and a script to link all the individual programs together. There are four major components of the package as give below:

**Super MC}: (the initial condition generator)**
This code can generate fluctuating initial conditions according to Monte-Carlo Glauber model and Monte-Carlo Kharzeev-Levin-Nardi (KLN) "gluon saturation" model. In this work, we have used the Monte-Carlo Glauber model initialization which assumes that the initial energy density in the transverse plane is proportional to the wounded nucleon density.

**VISH New: (viscous hydrodynamic simulator)**
Viscous Israel Stewart hydrodynamics is a (2+1)-d viscous hydrodynamic simulation for relativistic heavy-ion collisions. It solves the equation of motion for second order viscous hydrodynamics ("Israel-Stewart equations") with a given equation of state (EOS). VISHNew supports three versions of the lattice-based equation of state, s95p-v0-PCE, s95p-v1, and s95p-v1-PCE; all applying different implementations of partial chemical equilibrium in the hadronic phase. We have used s95p-v1-PCE for determining the pressure in the liquid, by first solving for the local energy density and velocity of the fluid cell.

**iSS: (a particle sampler)**
iSS is an "event generator" which generates a complete collision event of emitted hadrons, similar to the events created in the experiment, from Cooper-Frye freeze-out procedure. Finally:

**UrQMD: (afterburner)**
A hadron cascade simulator ideally suited for the description of the dynamics of a system of hadrons, both in and out of equilibrium.

**Solving Maxwell's Equation**
The electromagnetic fields generated by a point-like charge moving in the +z-direction with velocity (**v**) can be governed by the following wave equations [24].

$$\nabla^2 \vec{B} - \partial_t^2 \vec{B} - \sigma \partial_t \vec{B} = -\vec{\nabla} \times \vec{J}_{\text{ext}} \quad (4.1)$$

$$\nabla^2 \vec{E} - \sigma_t^2 \vec{E} - \sigma \partial_t \vec{E} = \frac{1}{\epsilon} \vec{\nabla} \rho_{\text{ext}} + \partial_t \vec{J}_{\text{ext}} \quad (4.2)$$

Where the external charge and current sources for the electromagnetic fields [24] which are a function of the rapidity of those flying and interacting particles are then given by:

$$\rho_{\text{ext}}(\vec{x}_\perp, \eta_s) = \rho_{\text{ext}}^+(\vec{x}_\perp, \eta_s) + \rho_{\text{ext}}^-(\vec{x}_\perp, \eta_s) \quad (4.3)$$

$$\vec{J}_{\text{ext}}(\vec{x}_\perp, \eta_s) = \vec{J}_{\text{ext}}^+(\vec{x}_\perp, \eta_s) + \vec{J}_{\text{ext}}^-(\vec{x}_\perp, \eta_s) \quad (4.4)$$

Solving the above wave equations happens to be easy since we have considered the electrical conductivity of the QGP to be constant, but making it time varying, which in reality is, because of its dependence on the temperature of the system would make it very hard and is what we anticipate to work on next. So we took four different (σ) values σ = 0.023 fm$^{-1}$, sigma = 0.4 fm$^{-1}$, sigma = 0.66 fm$^{-1}$ and sigma = 1.1 fm$^{-1}$ to have a chance of analyzing its effect. And the numerical code for calculating the electromagnetic fields is same as in [24] which can be found at https://github.com/chunshen1987/Heavy-ion_EM_fields}. For the evolution of the relativistic viscous hydro-dynamics, we choose the s95p-v1-PCE equation of state from [26]. After having the field evolution known the next imediate step is to work on what drift velocity this field causes and this can be done by solving the force balance equation.

**The Equation of Motion**
The drift velocity at each point on the freeze-out surface from the electromagnetic field evolution can be calculated after solving the force balance equation [24, 27]:

$$m\frac{d\vec{v}_{\text{drift}}^{\text{lrf}}}{dt} = q\vec{v}_{\text{drift}}^{\text{lrf}} \times \vec{B}^{\text{lrf}} + q\vec{E}^{\text{lrf}} - \mu m \vec{v}_{\text{drift}}^{\text{lrf}} = 0 \quad (4.5)$$

This gives us the local velocity due to electromagnetic effects. The last term in (4.5) describes the drag force on a fluid element with mass m on which some external (in this case electromagnetic) force is being exerted, with μ the drag coefficient. Even-though the drag coefficient still is not clear, at present its value is known precisely only for heavy quarks in N = 4 SYM theory; following [26, 27].

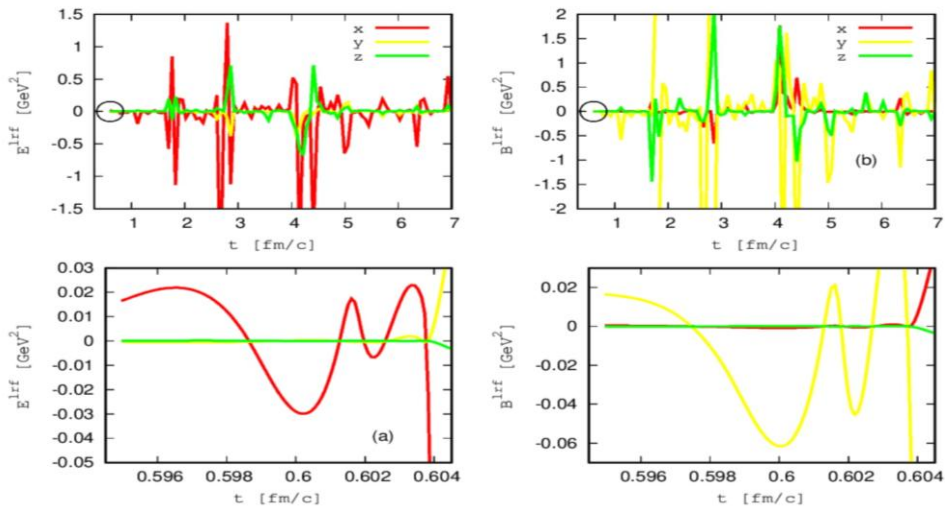

FIG. 3: (Color online) Illustrates evolution of the three components of the electric and magnetic fields in the local fluid rest frame at points on the freeze-out surface created in relativistic Pb+Pb collision with collision energy of $\sqrt{s} = 13$ A TeV. The calculations were made at centrality ranging from 20-30 % (impact parameters in the range 6.24 fm < b < 9.05 fm) and $\eta_s$ ranging from -7.9 to 7.9 where z is the reaction axis while x and y represents the transverse and longitudinal directions respectively. Figures in the right row illustrates evolution of the three components of the magnetic fields in the local fluid rest frame at points on the freeze-out surface created in relativistic Pb+Pb collision with collision energy of $\sqrt{s} = 13$ ATeV, and the three components of the magnetic fields evolution during the very early propagation time in the fluid rest frame at a fixed space time rapidity of $\eta_s = 0$. The top left column figure illustrates the evolution of the three components of the magnetic fields in the local fluid rest frame at points on the freeze-out surface and the bottom one dictates the three components of the electric fields versus the local time in the fluid rest frame at a fixed space time rapidity of $\eta_s = 0$.

We investigate the force balance equation from which we get the drift velocity $v^{\text{lrf}}$ in every fluid cell along the freeze-out surface for different values of q on the equation. We have considered charges of ±1, ± 2, ± 1/3 and ± 2/3. After having the drift velocity calculated, we added it with the usual hydrodynamic flow velocity; then after, we fed it back to the framework in order to find out the particles velocity increment.

| Percentage increase of flow harmonics at $\sqrt{s} = 13$ A TeV | | | | | | | | |
|---|---|---|---|---|---|---|---|---|
| No. | Pt (GeV) | Neutron | | | | Omega | | | |
| | | $\Delta V_1$ | $\Delta V_2$ | $\Delta V_3$ | $\Delta V_4$ | $\Delta V_1$ | $\Delta V_2$ | $\Delta V_3$ | $\Delta V_4$ |
| 1 | 0.007 | -1.71 | -1.01 | -0.04 | 0.00025 | -69.3 | -1.1 | -0.06 | 0.0001 |
| 2 | 0.093 | -1.75 | -1.00 | -1.48 | 0.23 | -70.5 | -1.1 | -1.5 | 0.18 |
| 3 | 0.038 | -1.78 | -1.01 | -1.87 | -1.33 | -69.5 | -1.1 | -1.8 | -0.9 |
| 4 | 0.282 | -1.83 | -1.04 | -1.91 | -1.1 | -66.3 | -1.2 | -1.9 | -0.59 |
| 5 | 0.175 | -1.93 | -1.17 | -1.95 | -0.88 | -61.0 | -1.6 | -1.97 | -0.35 |
| 6 | 0.582 | -2.1 | -2.2 | -2.1 | -0.5 | -56.0628 | 10.3 | -2.2 | 0.26 |
| 7 | 0.417 | -2.3 | 1.3 | -2.45 | 0.3 | -58.7 | 0.4 | -2.9 | 8.2 |
| 8 | 0.778 | -2.5 | 0.26 | -3.76 | -11.4 | -159.0 | 0.092 | -6.9 | -2.2 |
| 9 | 1.010 | -3.3 | 0.04 | -1.9 | -1.9 | 23.6 | -0.059 | 3.5 | -1.33 |
| 10 | 1.281 | -3.2 | -0.1 | 1.3 | -1.24 | 2.95 | -0.19 | 0.16 | -0.99 |
| 11 | 1.598 | -3 | -0.3 | -0.4 | -0.9 | -0.307 | -0.33 | -0.8 | -0.75 |
| 12 | 1.971 | 0.3 | -0.45 | -1.2 | -0.6 | -21.9 | -0.49 | -1.5 | -0.49 |
| 13 | 2.416 | -4.3 | -0.6 | -2.0 | -0.3 | -2.87 | -0.68 | -2.27 | -0.14 |
| 14 | 2.964 | -1.2 | -0.9 | -3.0 | 0.2 | 0.22 | -0.98 | -3.28 | 0.4 |
| 15 | 3.694 | 5.9 | -1.5 | -4.8 | 1.2 | 7.95 | -1.59 | -5.18 | 1.37 |

TABLE III: The percentage increase of flow harmonics due to the electromagnetic force evolution created in relativistic Pb+Pb collision with collision energy of $\sqrt{s} = 13$ A TeV. The change in elliptic flow ($\Delta V_2$) is percentage increase was calculated as the difference between $V_2$ Theory and $V_2$ with EM divided by $V_2$ Theory and the the whole resut was multiplied by 100 percent.

## RESULTS AND DISCUSSIONS

This work focused on including the electromagnetic field evolution calculation on the well known iEBE-VISHNU code package for relativistic heavy-ion collisions [28]. We studied the percentage increase of elliptic flow of identified particles due to the electromagnetic force evolution created in relativistic Pb+Pb collision with collision energy of $\sqrt{s}$ = 2.76, 5.02, 8.16 and 16 A TeV using similar model set up as [24]. The results show that the electromagnetic force affects the flow harmonics with different magnitude as explained below. In heavy ion collision the magnetic field is expected to be solenoidal fields from those flying charges. In line with this, during the beginning of the collision time, the dominant magnetic field is from the spectators, yet the participants of both incoming projectile ions contribute to the evolution of the electric and magnetic field created. In the next two sections we shall discuss on the evolution of the electromagnetic field and its effects on observables.

**The Electromagnetic field evolution**

Fig-2 is an illustration of the evolution of the electromagnetic field created in relativistic heavy-ion collision after a Pb+Pb collision with 20-30 centrality (impact parameters in the range 6.24 fm < b < 9.05 fm and with a collision energy $\sqrt{s}$= 2.76, 5.02, 8.16 and 13 ATeV.

| No. | Pt (GeV) | Neutron | | | | Omega | | | |
|---|---|---|---|---|---|---|---|---|---|
| | | $\Delta V_1$ | $\Delta V_2$ | $\Delta V_3$ | $\Delta V_4$ | $\Delta V_1$ | $\Delta V_2$ | $\Delta V_3$ | $\Delta V_4$ |
| 1 | 0.007 | -1.72 | -1.01 | -0.04 | 0.0002 | -69.3 | -1.1 | -0.1 | 0.0001 |
| 2 | 0.093 | -1.8 | -1.0 | -1.5 | 0.23 | -70.5 | -1.1 | -1.5 | 0.2 |
| 3 | 0.038 | -1.8 | -1.0 | -1.9 | -1.33 | -69.5 | -1.1 | -1.8 | -0.8 |
| 4 | 0.282 | -1.8 | -1.04 | -1.9 | -1.05 | -66.3 | -1.2 | -1.9 | -0.6 |
| 5 | 0.175 | -1.9 | -1.17 | -1.95 | -0.9 | -61.0 | -1.6 | -1.9 | -0.35 |
| 6 | 0.582 | -2.1 | -2.2 | -2.1 | -0.5 | -56.0 | 10.3 | -2.2 | 0.3 |
| 7 | 0.417 | -2.3 | 1.3 | -2.45 | 0.3 | -58.7 | 0.4 | -2.9 | 8.2 |
| 8 | 0.778 | -2.5 | 0.26 | -3.8 | -11.4 | -159 | 0.1 | -6.9 | -2.2 |
| 9 | 1.010 | -3.32 | 0.04 | -182.0 | -1.9 | 23.6 | -0.06 | 3.5 | -1.3 |
| 10 | 1.281 | -3.25 | -0.1 | 1.3 | -1.2 | 2.9 | -0.2 | 0.16 | -0.99 |
| 11 | 1.598 | -3.0 | -0.3 | -0.4 | -0.9 | -0.3 | -0.3 | -0.8 | -0.8 |
| 12 | 1.971 | 0.27 | -0.4 | -1.2 | -0.6 | -21.9 | -0.5 | -1.5 | -0.5 |
| 13 | 2.416 | -4.3 | -0.63 | -2.0 | -0.3 | -2.9 | -0.7 | -2.3 | -0.14 |
| 14 | 2.964 | -1.2 | -0.9 | -3.0 | 0.24 | 0.2 | -1.0 | -3.3 | 0.4 |
| 15 | 3.694 | 5.9 | -1.5 | -4.85 | 1.2 | 7.9 | -1.6 | -5.2 | 1.4 |

Percentage increase of flow harmonics at $\sqrt{s} = 8.16$ A TeV

TABLE II: The percentage increase of flow harmonics due to the electromagnetic force evolution created in relativistic Pb+Pb collision with collision energy of $\sqrt{s} = 8.16$ A TeV. The change in elliptic flow ($\Delta V_2$) is percentage increase was calculated as the difference between $V_2$ Theory and $V_2$ with EM divided by $V_2$ Theory and the the whole resut was multiplied by 100 percent.

The three components of the electric field in the local fluid rest frame at points on the freeze-out surface are shown on Fig-2. As explained in the previous sections the total electric field comes from Coulomb field of the spectators and the plasma it self, the Faraday and also the Lorentz field from the moving charges. The created electric field seen above and below of the reaction axis is quite different and this is expected due to the coulomb electric field created by the positively charged spectator particles at the beginning of the collision. This field created from the spectators is the reason for the current created in the plasma. The electric field in the z-direction is shown to be smaller than that of the two axes. This field created from the spectators is the very reason for the current created in the plasma. The z-direction being the axis of collision, the electric field in its direction is not as wide and broad as that of the other two axis which are a direct evidence of the two projectiles creating fields in opposite direction and the complex multi-directional expansion of the system puts the created fields down as the system evolves. The coulomb force between the particles in the plasma creates an electric field which is against the coulomb electric field. The decreasing of the electric field in the reaction plane in-turn drops the magnetic field causing Faraday electric field to evolve. The electric force being multi directional contributes to the even flow harmonics.

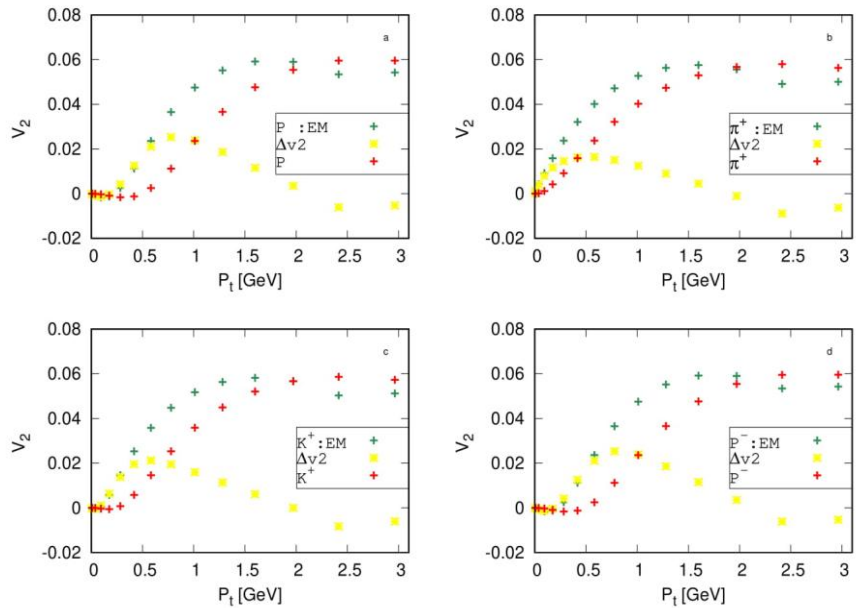

FIG. 4: (Color online) The effect of electromagnetic force on the change in directed flow $\Delta V_1$, change in elliptic flow $\Delta V_2$, change in triangular flow $\Delta V_3$ and change in rectangular flow $\Delta V_4$ for some identified particles. The change in even flow harmonics coefficients is higher than that of the odds.

One can note that the magnetic field eB reaches the value about 0.4 GeV$^2$ which joins the manifestation of Chiral Magnetic Effect (CME), and this magnetic field has the ability to affect chirality at the early stage of the QGP evolution. The electromagnetic field created in such relativistic energy is strong enough to create changes for the behavior of every light or heavy particles making it so complex to trace its effect clearly.

**The elliptic flow**

Table-I presents how the percentage increase in elliptic flow of the four particles kaon plus, pion plus, pion minus and proton vary. The elliptic flow of proton is suppressed up to an elliptic flow value of nearly 2.37% of the initial value. At lower momentum they get suppressed down to 1 % of its initial value. It is different for pions; at lower momentum the created field raised the flow up to 2.7%. The effect of the electromagnetic evolution is almost same for the considered particles at higher momentum. As-far-as the system evolution is concerned the flow harmonics values gets suppressed or rouse is non uniform fashion through out the evolution. The percentage increases in elliptic flow of Pion-plus and Pion-minus are exactly the same, dictating how mass plays a dominant role in affecting the bend of flow of particles due to the electromagnetic field evolution.

In Table II and III the momentum-dependent flow harmonics coefficients for Neutron and Omega are illustrated. The elliptic flow for the ideal case is compared to the case with the electromagnetic field. The electromagnetic field acts on the evolution of flow that leads to the reduction of elliptic flow of these positively charged particles dominantly. The flow harmonics percentage increase of both neutron and omega increase as the collision energy is raised. At lower momentum both neutron and omega get a percentage increase less than 2% from its initial value. At higher momentum the directed flow is raised up to 6% for neutron and 8 % for omega. The created electromagnetic field raised the directed flow in non uniform fashion up to 70 %. Yet, the elliptic flow at 13 A TeV is raised to percentage increase of 10 % which is just 2.7% at 2.76 A TeV.

As-far-as the system evolution is concerned, the flow harmonics values get suppressed or rouse in non uniform fashion through out the evolution. Inorder to explain the effects caused by the electromagnetic evolutions better, every possible contributors like the drag force, electrical conductivity and gluon charge density shall be explained well.

As it is stated in Fig-4, the electromagnetic field effect on $v_2$ is small in magnitude for the positively charged and lighter particles. In addition to this, Fig-4 dictates that the elliptic flow of particles get raised at the beginning of the evolution and suppressed latter. Moreover, proton and anti-proton get similar effect.

A hand full of other groups have studied the effect of electromagnetic field evolution on the flow of particles in relativistic heavy ion collision. They all have used different approaches making some crude assumptions to tackle the problem as we just did. All the studies [24, 33, 34] came up to similar understanding as the electromagnetic evolution has an effect on the bending of particles flow. Moreover, the result from [34] on flow harmonics changes of heavier particles agrees with ours.

**Other Identified Particles**

As results presented above proved, the flow harmonics of particles is influenced by the electromagnetic field evolution. Here let's see how other particles are influenced too. The elliptic flows of heavier elementary particles are influenced in greater magnitude by the fluid evolution. The plot dictates that the change in directed flow is so small that even one can say, the side push from the field is almost insignificant. Even in these very small changes proton, neutron and kaons get pushed aside with higher magnitude than that of pions. This tells us that, the heavier the particle is the higher side pushes it gets.

One can see from Fig-5, the electromagnetic field effect on Δv1 is small in magnitude. this can lead us to ask that: how those electric fields (the coulomb, Faraday, Lorentz, and the field from the plasma) do not cause a bigger change on the directed flow of particles? The answer would rather be simple, they cancel out each other and get extremely low so that the field in the dilute system manages to be dominant multi-directionality. Because of this the contribution to the odd harmonics fall to be very small and a significant contribution to the even harmonics is clearly seen. From these one can come up to the decision that, the side-way electric fields created by the Faraday and Coulomb has almost canceled the electric Lorentz field created in the fluid system still having very small contribution to the directed flow and this is on the collision axis.

To the contrary of the results from [24], we have found that the electromagnetic field evolution does depend not much on the type of charge the particles acquire but on the mass of particle considered. This is clearly seen on the table I, II, III and Fig-5 that a particle and its anti particle gets a side push from the electromagnetic field evolution with very close magnitudes. In the force balance calculation, we found the drift velocity for particles with charge 1, -1, 2 and -2 as in [24]. Diversifying the vast charge consideration will await us for future exploration of the field influence on any particle found anywhere.

Finally, Fig-5 dictates that particle lambda and Xi gets larger directed flow change and this tells us that lambda and xi gets a better kick off than even proton; same as found in Ref. [35]. This tells us once again the dependence of field effect on the mass of created particles reminding as both contains strange and (or) charm quarks. A hand full of other groups have studied the effect of electromagnetic field evolution on the flow of particles in relativistic heavy ion collision. They all have used different approaches making some crude assumptions to tackle the problem as we just did. Our model set-up is same as that given in [34]. All the studies [33, 34, 36] came up to similar understanding as the electromagnetic evolutions has an effect on the bending of particles. Our result of heavier particles getting better initial push agrees with [34].

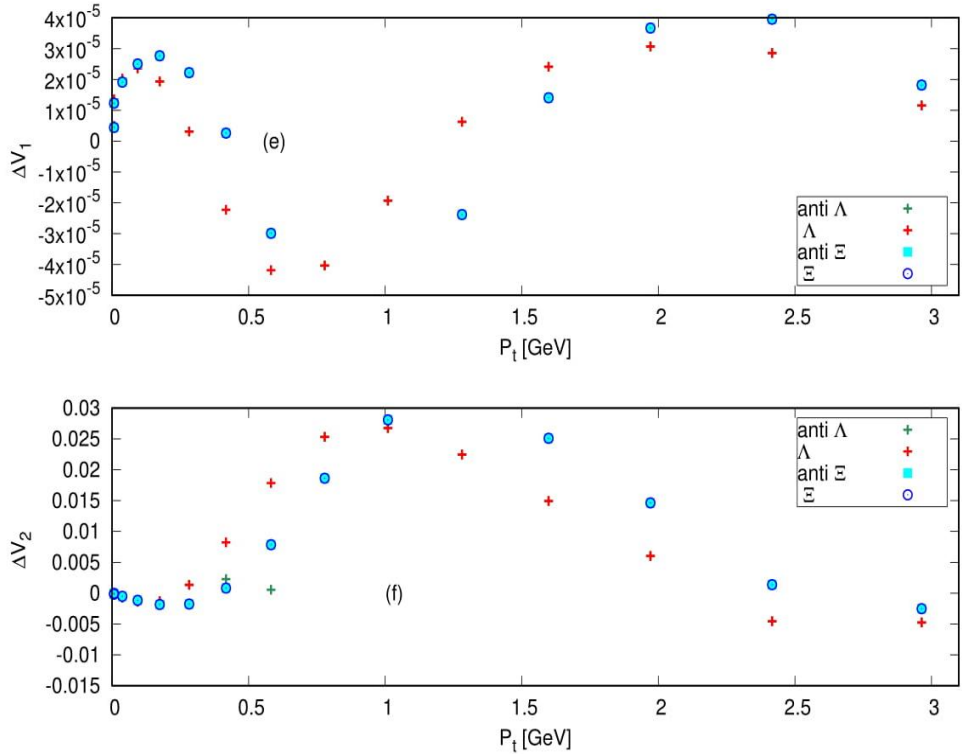

FIG. 5: (Color online) Illustration of the effect of electromagnetic force on the change in directed flow $\Delta V_1$ and change in elliptic flow $\Delta V_2$ for two identified particles and their anti−particles. The change in even flow harmonics coefficients is higher than that of the odds. plots (b) and (d) dictates that, the change in elliptic flow $\Delta V_2$ is in higher order than the change in directed flow $\Delta V_1$

## SUMMARY AND CONCLUSIONS

We have employed the iEBE-VISHNU frame work to investigate the generation and evolution of the electromagnetic fields in relativistic heavy-ion collisions. The evolution of the electromagnetic field created after a Pb+Pb relativistic collision with 20-30 centrality and collision energy of $\sqrt{s}$ = 2.76, 5.02, 8.16 and 13 ATeV was explored. The spatial cognition of the electromagnetic field was surveyed and a very inconsistent dispersion is recovered. We have also examined the time evolution of the fields at the early-stage of the development. We uncovered that the residues give considerable contribution to the fields during the early-stages evolution. A study of the effect of this external electromagnetic field on elliptic flow of particles was investigated. Both the ideal and the electromagnetic field cases were initialized equally and the transverse profile was taken from a Monte-Carlo Glauber model initialization calculation.

As a result, a maximum of ± 2.7 % increase in elliptic flow is observed at collision energy of 2.76 ATeV and percentage increase of 10 % at collision energy of 8.16 ATeV is found. Protons being massive than pions, have got affected by the field in the early evolution time. The elliptic flow coefficient is affected to the order of a thousand than the directed flow, and this happens because of the nullification of the electric fields in the x and y direction having a slight field in the reaction axis which causes the very small directed flow change observed. Besides, we find out that heavier particles like λ and Ξ and get larger directed flow change than the other particles like π, kaons, and even proton. Their building blocks being strange and/or charm quarks played a great role to let them behave this way. Mass is the dominant factor than charges which is also observed as particles and their anti-particles get washed aside by the field in similar fashion.

To conclude, the present study shows that beside the inclusion of electromagnetic field, the increase in collision energy increases the elliptic flow of particles in non uniform fashion through out the evolution. At last, further study is needed to establish a better understanding on the electromagnetic field evolution and its effects on the created system by softening many of the crude assumptions we made and keeping the functionality of parameters.

## ACKNOWLEDGMENTS
We would like to thank Chun Shen for the briefing discussion on the codes we have used for the frame work, and we would like to forward our gratitude to Scott Pratt of Michigan State University for the valuable discussions we have had at the start of the work.